\documentclass[aps,floatfix,superscriptaddress,reprint,10pt,prl]{revtex4-1}
\usepackage{bbm}
\usepackage{bm}
\usepackage{amsmath}
\usepackage{amssymb}
\usepackage{empheq}
\usepackage{graphicx}
\usepackage{mathrsfs}
\usepackage{amsfonts}
\usepackage{amsthm}
\usepackage{color}
\usepackage{bigints}
\usepackage{txfonts}
\usepackage{hyperref}
\hypersetup{
     unicode=false,          % non-Latin characters in Acrobat?s bookmarks
     pdftoolbar=true,        % show Acrobat?s toolbar?
     pdfmenubar=true,        % show Acrobat?s menu?
     pdffitwindow=false,     % window fit to page when opened
     pdfstartview={FitH},    % fits the width of the page to the window
     pdftitle={My title},    % title
     pdfauthor={Author},     % author
     pdfsubject={Subject},   % subject of the document
     pdfcreator={Creator},   % creator of the document
     pdfproducer={Producer}, % producer of the document
     pdfkeywords={keyword1} {key2} {key3}, % list of keywords
     pdfnewwindow=true,      % links in new PDF window
     colorlinks=false,       % false: boxed links; true: colored links
     linkcolor=red,          % color of internal links (change box color with linkbordercolor)
     citecolor=green,        % color of links to bibliography
     filecolor=magenta,      % color of file links
     urlcolor=cyan           % color of external links
}
\makeatletter \tolerance = 10000 \tolerance = 10000
\usepackage{color}

\makeatother
\begin{document}
\title{Unusual Metal to Marginal-Metal Transition in
Two-Dimensional Ferromagnetic Electron Gases}
%\title{Unconventional Quantum Phase Transitions in
%Two-Dimensional Ferromagnetic Electron Gases with
%Rashba Spin-Orbit Interactions: Emergence of Marginal Metals
%between Diffusive Metals and Anderson Insulators}

\author{Weiwei Chen}
\affiliation{Hefei National Laboratory for Physical
Sciences at the Microscale $\&$ Synergetic Innovation
Center of Quantum Information and Quantum Physics,
University of Science and Technology of China, Hefei, Anhui 230026, China}
\author{C. Wang}
\email[Corresponding author: ]{cwangad@connect.ust.hk}
\affiliation{School of Electronic Science and Engineering
and State Key Laboratory of Electronic Thin Films and
Integrated Devices, University of Electronic Science and Technology
of China, Chengdu 610054, China}
\author{Qinwei Shi}
\affiliation{Hefei National Laboratory for Physical
Sciences at the Microscale $\&$ Synergetic Innovation
Center of Quantum Information and Quantum Physics,
University of Science and Technology of China, Hefei, Anhui 230026, China}
\author{Qunxiang Li}
\affiliation{Hefei National Laboratory for Physical
Sciences at the Microscale $\&$ Synergetic Innovation
Center of Quantum Information and Quantum Physics,
University of Science and Technology of China, Hefei, Anhui 230026, China}
\author{X. R. Wang}
\email[Corresponding author: ]{phxwan@ust.hk}
\affiliation{Department of Physics, The Hong Kong University of Science and Technology (HKUST),
Clear Water Bay, Kowloon, Hong Kong}
\affiliation{HKUST Shenzhen Research Institute, Shenzhen 518057, China}

\date{\today}
\begin{abstract}
Two-dimensional ferromagnetic electron gases subject to random scalar potentials
and Rashba spin-orbit interactions exhibit a striking quantum criticality.
As disorder strength $W$ increases, the systems undergo a transition from
a normal diffusive metal consisting of extended states to a marginal metal
consisting of critical states at a critical disorder $W_{c,1}$.
Further increase of $W$, another transition from the marginal metal to an
insulator occurs at $W_{c,2}$. Through highly accurate numerical procedures based
on the recursive Green's function method and the exact diagonalization, we elucidate
the nature of the quantum criticality and the properties of the pertinent states.
The intrinsic conductances follow an unorthodox single-parameter scaling law:
They collapse onto two branches of curves corresponding to diffusive metal phase and
insulating phase with correlation lengths diverging exponentially
as $\xi\propto\exp[\alpha/\sqrt{|W-W_c|}]$ near transition points.
Finite-size analysis of inverse participation ratios reveals that the states
within the critical regime $[W_{c,1},W_{c,2}]$ are fractals of a universal
fractal dimension $D=1.90\pm0.02$ while those in metallic (insulating)
regime spread over the whole system (localize) with $D=2$ ($D=0$).
A phase diagram in the parameter space illuminates the occurrence and
evolution of diffusive metals, marginal metals, and the Anderson insulators.
\end{abstract}
\maketitle

Anderson localization is a long-lasting fundamental concept in condensed matter physics
\cite{pwanderson1,sjohn1,dswiersma1,mstorzer1,aachabanov1,psheng1,hhu1,tschwartz1,cwang1}
and keeps bringing us surprising, especially
in its critical dimensionality of two \cite{palee1,bkramer1,fevers1}.
In the early time, the orthodox view is the absence of diffusion of an
initially localized wave packet at an arbitrary weak disorder in one- and
two-dimensional electron gases (2DEGs) while metallic states and Anderson
localization transitions (ALTs) can occur in 3D \cite{eabrahams1,dfriedan1}.
Later more careful renormalization group calculations \cite{shikami1}
and numerical simulations \cite{snevangelou1,rmerkt1,yasada1,pmarkos1,gorso1},
together with experiments \cite{bkramer1}, show that intrinsic degrees of
freedom can alter the results in 2D: Half-integer spin particle systems can
also support ALTs when the spin rotational symmetry is broken through
spin-orbit interactions (SOIs), regardless whether the time reversal symmetry
is preserved (symplectic class) \cite{shikami1,snevangelou1,rmerkt1,
yasada1,pmarkos1,gorso1} or not (unitary class) \cite{cwang2,ysu1,cwang3}.
The most unquestionable examples would be quantum Hall effects of both
non-interacting \cite{djthouless1,ammpruisken1,bhuckestein2} and interacting
\cite{dctsui1,rblaughlin1} 2DEGs in strong perpendicular magnetic fields.
On the other hand, all states of disordered non-interacting integer-spin
particle systems must be localized \cite{rsepehrinia1,ysu2}.
\par

\begin{figure}
\includegraphics[width=0.45\textwidth]{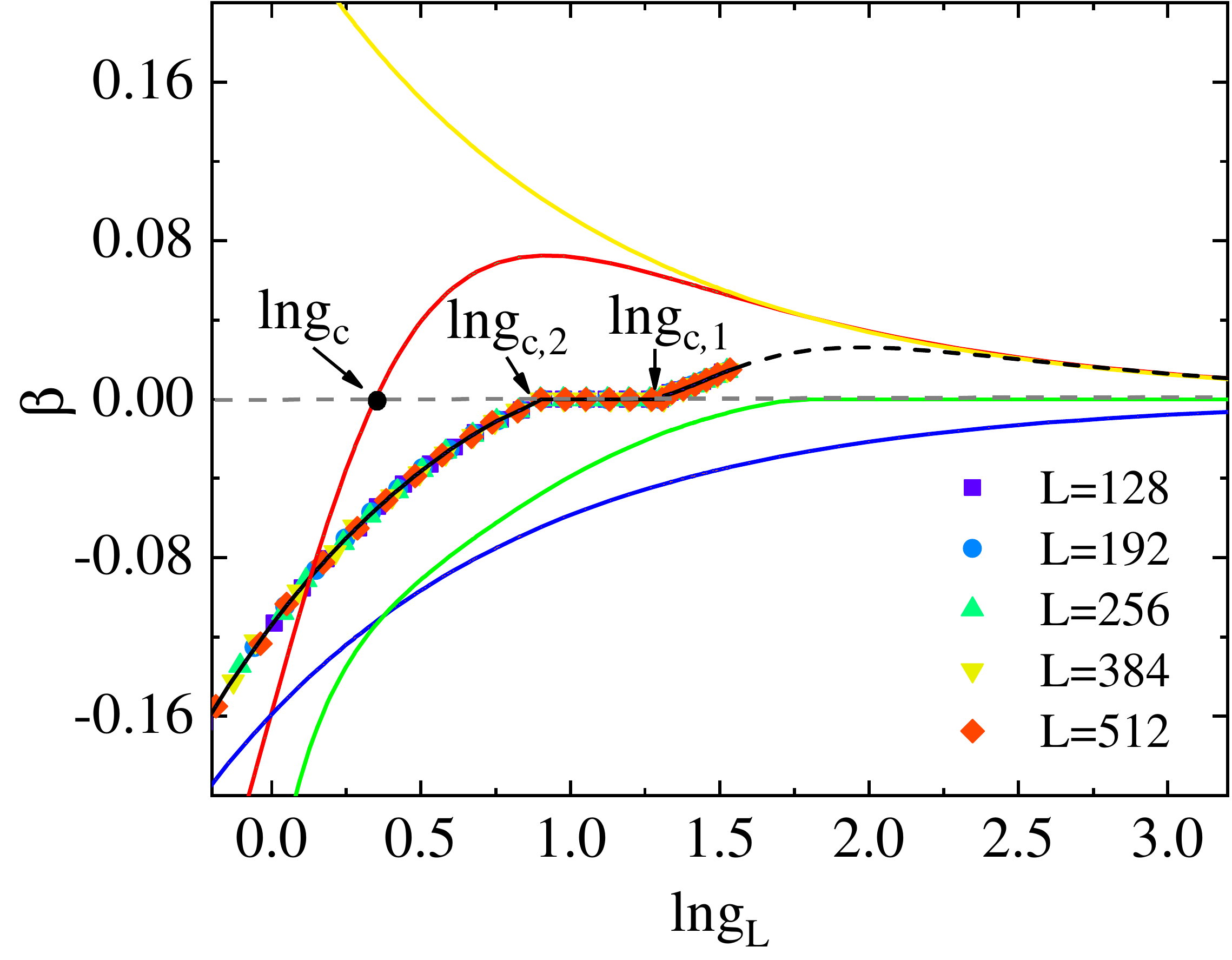}
\caption{$\beta(\ln g_L)$ of various 2D materials.
Black solid line stands for the numerical results (symbols in Fig.~\ref{fig2}(a)
for different system sizes) presented in this work, which displays
the coexistence of DM ($\beta>0$), MM ($\beta=0$), and AI ($\beta<0$).
While, black dash line is plotted according to an analytical formula \cite{note_beta,shikami1}.
For non-interacting Schr\"{o}dinger electrons, no delocalization-localization
transition is allowed for orthogonal class (blue line) \cite{bkramer1,fevers1},
where time-reversal and spin-rotation symmetries are preserved.
For symplectic class \cite{snevangelou1,rmerkt1,yasada1,pmarkos1,gorso1,yasada2},
there is one $\beta=0$ point (red line), corresponding to an unstable fixed point in renormalization group flow.
The KT transition (green line) from a band of critical
states to localized states can also exist \cite{cwang2}. While for the Dirac
Hamiltonians, numerical simulations suggest all states are extended in
single-valley graphenes \cite{jhbardarson1} (yellow line) while KT transitions
are also allowed if random fluxes \cite{xcxie1} or intervalley scatterings
\cite{yyzhang1,zxu1} exist.}
\label{fig1}
\end{figure}

However, recent numerical studies \cite{xcxie1,yavishai1,gxiong1,cwang2,
ysu1,cwang3,jbang1} showed that the current understanding of ALTs in
non-interacting 2DEGs is far from completed when SOIs are involved.
For example, in contrast to the predictions based on the nonlinear
$\sigma$ model that claims only localized states are allowed in
the unitary ensemble \cite{shikami1}, a band
of extended states together with an ALT or critical
states usually accompanied by a Kosterlitz-Thouless (KT) transition
could exist in 2DEGs without time-reversal symmetry, depending on the
form of SOIs and the strength of magnetic field \cite{cwang2,ysu1,cwang3}.
In this work, we observe an anomalous phase transition from a normal metal
to a {\it marginal metal} \cite{yyzhang1}, consisting of a band of metallic
critical states, in a ferromagnetic 2DEG on a square lattice subject to a Rashba SOI
and random on-site potentials.
The statements are supported by argumentations
based on two independent highly-accurate numerical approaches: the
finite-size scaling analysis of two-terminal conductances and the inverse
participation ratios (IPRs) analysis of wave functions obtained from
the exact diagonalization. The unaccustomed marginal metal (MM) phase
exists between a diffusive metal (DM) phase at weak disorders and an
Anderson insulator (AI) phase at strong disorders.
Scaling analyses of IPRs show that wave functions of states in the MMs
are of fractals of dimension $D=1.90\pm0.02$, a feature reminiscent of
a band of critical states in the random SU(2) model subject to strong
magnetic fields \cite{cwang2}.
\par

Our main result is the new (marginal metallic) phase
whose $\beta$-function
(symbols and black line) defined as $\beta(\ln g_L)=d\ln g_L/d \ln L$
describes an unconventional DM-MM-AI transition, as summarized in Fig.~\ref{fig1}.
In comparison, $\beta$ for other types of phase transitions in 2D
are also included (elucidate in the caption). Here $g_L$ and $L$
are dimensionless conductance and system size, respectively.
For a large (small) conductance, i.e., $g_L>g_{c,1}$
($g_L<g_{c,2}$), $\beta$ is positive (negative),
indicating a metallic (insulating) phase.
While between two critical conductances $g_{c,1}$ and $g_{c,2}$,
$\beta=0$ shows a MM phase in which conductances are through critical
states that are size independent. Different from an ALT at a fixed point
$g_{c}$ \cite{snevangelou1,rmerkt1,yasada1,pmarkos1,gorso1,ysu1,cwang3},
the MM phase between $[g_{c,1},g_{c,2}]$ is a fixed line
in which the system does not flow away when its size is scaled.
Furthermore, near both DM-MM and MM-AI transition points, correlation
lengths $\xi$ locating on the DM and AI sides, respectively, diverge with
disorder strength $W$ as $\xi(W)\propto \exp[\alpha/\sqrt{|W-W_c|}]$,
a similar finite-size scaling law in KT transitions
(green line) \cite{cwang2,yyzhang1,xcxie1}.
\par

Our model is a tight-binding Hamiltonian on a square lattice of size
$L^2$ with unit lattice constant,
\begin{equation}
\begin{gathered}
H=\sum_{i}c_{i}^{\dagger}\epsilon_{i}
c_{i}-\sum_{\langle i,j\rangle}c_{i}^{\dagger}R_{ij}c_{j},
\end{gathered}\label{hamiltonian1}
\end{equation}
where $c^\dagger_{i}=(c^\dagger_{i,\uparrow},c^\dagger_{i,\downarrow})$
and $c_{i}$ are, respectively, the single electron
creation and annihilation operators on site $i=(x_i,y_i)$
with $x_i,y_i$ being integers and $1\leq x_i,y_i \leq L$.
$\langle ij\rangle$ denotes $i$ and $j$ as the nearest-neighbor sites.
The first term stands for on-site energy:
\begin{equation}
\begin{gathered}
\epsilon_i=\epsilon_{0}\sigma_0-\Delta\sigma_z+V_i\sigma_0.
\end{gathered}\label{hamiltonian2}
\end{equation}
Here $\sigma_0$ and $\sigma_{x,y,z}$ are the 2-by-2 identity and
the Pauli matrices respectively. $\epsilon_{0}$ is a constant energy term.
$-\Delta\sigma_z$ introduces a ferromagnetic term that breaks the time-reversal
symmetry with $\Delta$ quantifying the mean-field exchange splitting \cite{nnagaosa1}.
Disorders are modelled by $V_i\sigma_0$ with uncorrelated random numbers $V_i$
following the normal distribution of zero mean and the variance of $W^2$.
Thus, $W$ measures the degree of randomness. A Rashba SOI is encoded in the hopping
matrices $R_{ij}$, parameterized by matrices $R_x$ and $R_y$ along the $x-$
and $y-$directions, respectively,
\begin{equation}
\begin{gathered}
R_x=\dfrac{1}{2}(t\sigma_0+i\tilde{\alpha}\sigma_y)
\quad\text{and}\quad
R_y=\dfrac{1}{2}(t\sigma_0-i\tilde{\alpha}\sigma_x).
\end{gathered}\label{hamiltonian3}
\end{equation}
where $t$ is the energy unit and $\tilde{\alpha}$ measures the strength of SOIs.
In the clean limit, model~\eqref{hamiltonian1} can be blocked diagonalized in the
momentum space as $H=\sum_{\bm{k}}c^\dagger_{\bm{k}}h(\bm{k})c_{\bm{k}}$ with
\begin{equation}
\begin{gathered}
h(\bm{k})=(\epsilon_0-\cos k_x-\cos k_y)\sigma_0-\Delta\sigma_z\\
+\tilde{\alpha}(\sin k_x\sigma_y-\sin k_y\sigma_x).
\end{gathered}\label{hamiltonian4}
\end{equation}
Hereafter we fix $\epsilon_0=2$ such that the effective
$\bm{k}\cdot\bm{p}$ Hamiltonian near the band edge reads
$\frac{\bm{p}^2}{2}+\tilde{\alpha}(\bm{p}\times\bm{\sigma})\cdot\hat{z}$.
This form of Hamiltonians has been widely employed
to enlighten the intrinsic and extrinsic mechanism of the
anomalous Hall effect, and possible physical realizations of
model~\eqref{hamiltonian1} include a large family of ferromagnetic
semiconductors such as GaAs and other III-V host materials \cite{nnagaosa1}.
\par

To investigate the localization properties of states of model \eqref{hamiltonian1}, we
employ the Landauer formula to calculate the dimensionless conductance of a disordered
sample between two clean semi-infinite leads at a given Fermi level $E$, $\tilde{g}_L=
\text{Tr}[TT^\dagger]$, where $T$ is the transmission matrix \cite{amackinnon1}.
To exclude the contribution from contact resistances, we define the
dimensionless conductance $g_L$ as, $1/g_L=1/\tilde{g}_L-1/N_c$, where $N_c$ is
the number of propagating modes in the leads at Fermi energy $E$ \cite{sdatta1}.
The determination of quantum phase transitions is based on the following criteria:
(1) For the Fermi level in the DM (AI) phase, $dg_L/dL$ is positive (negative),
while in the MM phase, $g_L$ is independent of $L$.
(2) In the vicinity of phase transition points, there exists the one-parameter
scaling hypothesis \cite{eabrahams1} such that $g_L$ of different system
sizes merge into a universal smooth scaling function $f(x)$, i.e.,
\begin{equation}
\begin{gathered}
g_L=f(L/\xi),
\end{gathered}\label{scaling}
\end{equation}
with the correlation length $\xi$ diverging at the transition points.
\par

\begin{figure}[htbp]
\includegraphics[width=0.45\textwidth]{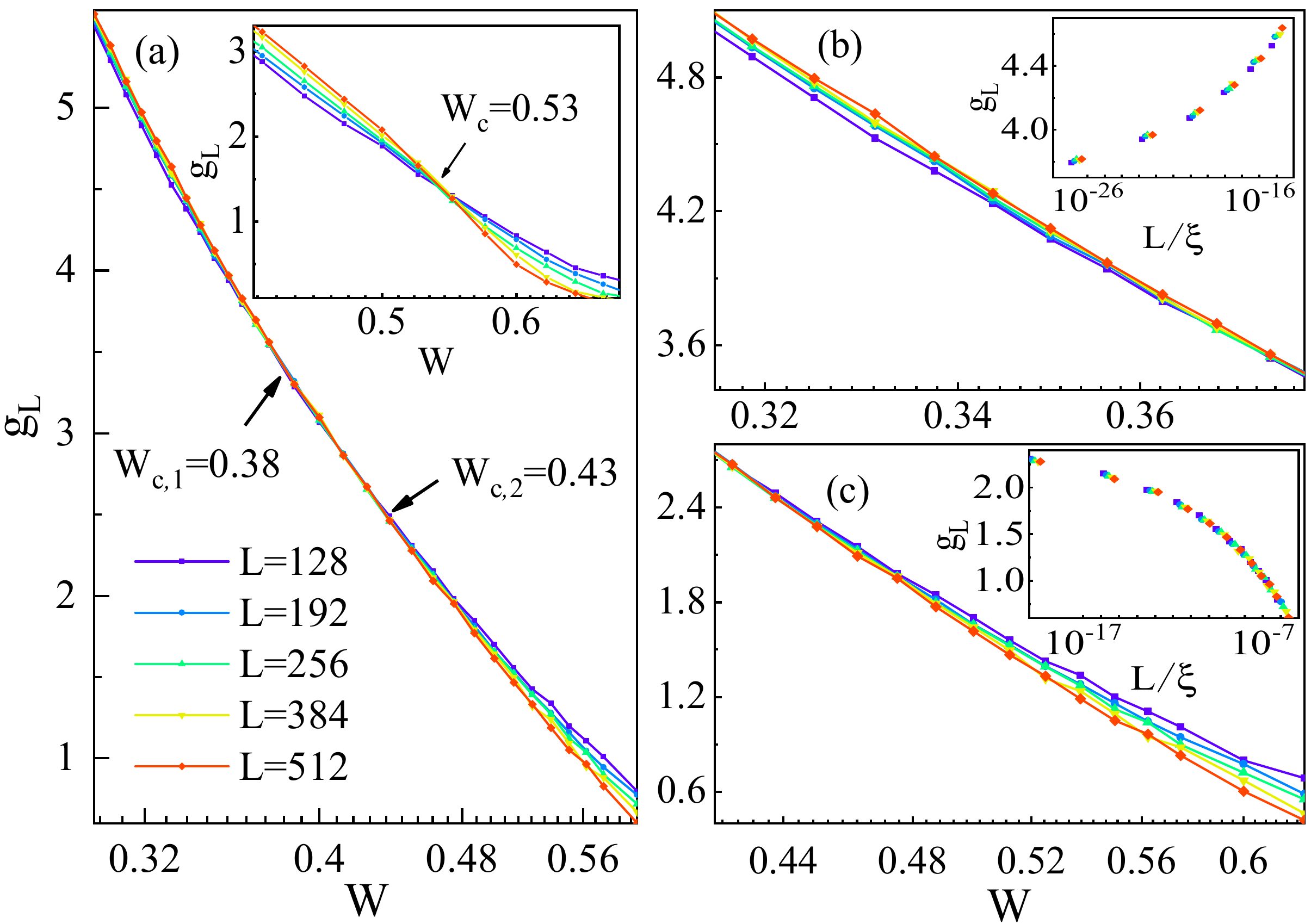}
\caption{(a) $g_L$ as a function of $W$ for $L=128$, 192,
256, 384, and 512 at $E=0.2$ and $\Delta=0.01$. Inset
is the same plot but for $\Delta=0$. (b) Enlargement
of the regime near $W_{c,1}$. Inset: Scaling function
$f(x=L/\xi)$ obtained by collapsing data for $W<W_{c,1}$
into a single curve. (c) Same as (b) but for the regime
near $W_{c,2}$.
}
\label{fig2}
\end{figure}

Two typical examples are shown in Fig.~\ref{fig2}(a) and its inset
that plot $g_L$ as a function of $W$ for $E=0.2$ (near band edge),
$\tilde{\alpha}=0.2$, and $\Delta=0.01$ and $0$ (inset), respectively.
Clearly, in the absence of the ferromagnetic coupling $\Delta=0$
when model~\eqref{hamiltonian1} belongs to the symplectic class,
all curves cross at a single point $W_c$.
$dg_L/dL$ is positive (negative) for $W$ smaller (larger) $W_c$.
These features are concrete evidence of an ALT at $W_c$.
Finite-size scaling analysis \cite{supp} shows that $g_L$ for
different sizes $L$ collapse to a single smooth scaling curve,
and $\xi$ diverges as $|E-E_c|^{-\nu}$ with $\nu=2.8\pm 0.2$,
consistent with previous estimates \cite{yasada1,pmarkos1,gorso1}.
\par

Strikingly, once systems enter the unitary class by turning on
the ferromagnetic coupling, say $\Delta=0.01$, we observe a MM phase
in the window of $W\in[W_{c,1},W_{c,2}]$ ($W_{c,1}=0.38\pm0.01$ and
$W_{c,2}=0.43\pm 0.01$) within which $dg_L/dL=0$ for all $L$, while
states for $W<W_{c,1}$ and $W>W_{c,2}$ are extended and localized,
respectively. Two phase transitions are evident in Figs.~\ref{fig2}
(b) and (c), which illustrate the enlargements near $W_{c,1}$
and $W_{c,2}$, respectively. The MM-AI transition at $W_{c,2}$,
in addition to the zero-plateau of $\beta$ function shown in
Fig.~\ref{fig1}, is highly evocative of the KT criticality arising
in another unitary ensemble with random SOIs \cite{cwang2}
and the graphene with random fluxes or long-range impurities
\cite{xcxie1,yyzhang1}. Nonetheless, the DM-MM transition at
$W_{c,1}$ from a band of extended states to a band of critical
states is highly nontrivial since both of them are of metallic
phases in the sense that their wave functions spread over the
whole lattice (illustrate later). To the best of our knowledge,
this kind of disorder-driven metal-metal transitions has never
been observed before in 2D materials, but in 3D semimetals
\cite{pgoswami1,kkobayashi1,broy1,czchen1,bfu1}.
\par

To substantiate the validity of one-parameter scaling hypothesis,
we show that all curves in Figs.~\ref{fig2}(b,c) collapse into two
smooth functions $f(x=L/\xi)$ shown in the insets of the figures
\cite{supp}, which offer direct verifications of quantum phase
transitions at $W_{c,1}$ and $W_{c,2}$. On the insulating side
and near the phase transition point $W_{c,2}$, the correlation
length is expected to diverge as $\xi\propto\exp[\alpha_{2}
/\sqrt{|W-W_{c,2}|}]$ with $\alpha_2=8.0\pm 0.8$, a fingerprint
of the KT transitions \cite{xcxie1,yyzhang1,cwang2}.
Differently, there are no reliable analytical and numerical
estimates for the divergence law near the DM-MM transition $W_{c,1}$.
Scaling analysis also suggests $\xi\propto\exp[\alpha_1/
\sqrt{|W-W_{c,1}|}]$ with the exponent $\alpha_1=9\pm 3$.
Besides, a power-law divergence of correlation lengths for ALTs,
i.e., $\xi\propto|W-W_c|^{-\nu}$, also fit the numerical data with
$\nu=32$. However, the obtained $\nu$ is much larger than any known
critical exponents in disordered 2D systems \cite{fevers1,bkramer1}.
We thus argue that at the DM-MM transition the correlation lengths show
the same scaling behavior as those for KT transitions,
rather than the power-law divergence for ALTs.
\par

So far, we have provided one example of the DM-MM-AI
transition in model~\eqref{hamiltonian1}. Needless to say, many
questions arise, and, among them, the most important one may be
the proof of the universality of such a quantum phase transition.
In Supplemental Materials \cite{supp}, we show indications of
universality by substantiating emergences of the three phases at
different Fermi energies with the same divergence law of $\xi$,
i.e., $\xi\propto\exp[\alpha/\sqrt{|W-W_c|}]$ at critical points.
The same physics is observed if we choose $E$ as the scaling variable
at a fixed disorder. Furthermore, simulations for a different form
of disorders and a distinct SOI due to the Dresselhaus effect are
both in qualitative agreement with Fig.~\ref{fig2}.
All these features indicate that the MM phase prevails in
ferromagnetic 2DEGs with SOIs and favors the exponential divergence
of $\xi$ at critical points.
\par

Having established the universality of DM-MM-AI transitions,
we further consider the nature of wave functions in three phases,
especially the fractal structure of wave functions in MM phase.
Wave functions at an isolated critical point of an ALT or in the
critical band are known to have multifractal structures
characterizing by a set of anomalous dimensions measuring
how their moments scale with sizes \cite{fevers1,cwang2,yyzhang1}.
Among them, the fractal dimension plays a pre-eminent
role, which is related to the IPR defined as
\begin{equation}
\begin{gathered}
p_2(E,W)= \sum_{i} |\psi_i(E,W)|^4
\end{gathered}\label{ipr}
\end{equation}
with $\psi_i(E,W)$ being the renormalized wave functions of
energy $E$ and disorder $W$ at site $i$ for a specific realization.
For large enough systems, the average IPR scales with size $L$ as
\cite{xrwang1,mjanssen1,jhpixley1}
\begin{equation}
\begin{gathered}
p_2(E,W)\propto L^{-D}
\end{gathered}\label{ipr_scaling}
\end{equation}
with $D$ being the fractal dimension.
If the state is extended (localized), $D=d$ ($D=0$),
where $d=2$ is the spatial dimension.
While for a critical state, an anomalous scaling with $L$ is
expected, i.e., $D\in [0,d]$. Thus, we expect that states
in the MM phase have a universal fractal dimension such that
their wave functions occupy much sparse space like fractals.
\par

\begin{figure}[htbp]
\includegraphics[width=0.45\textwidth]{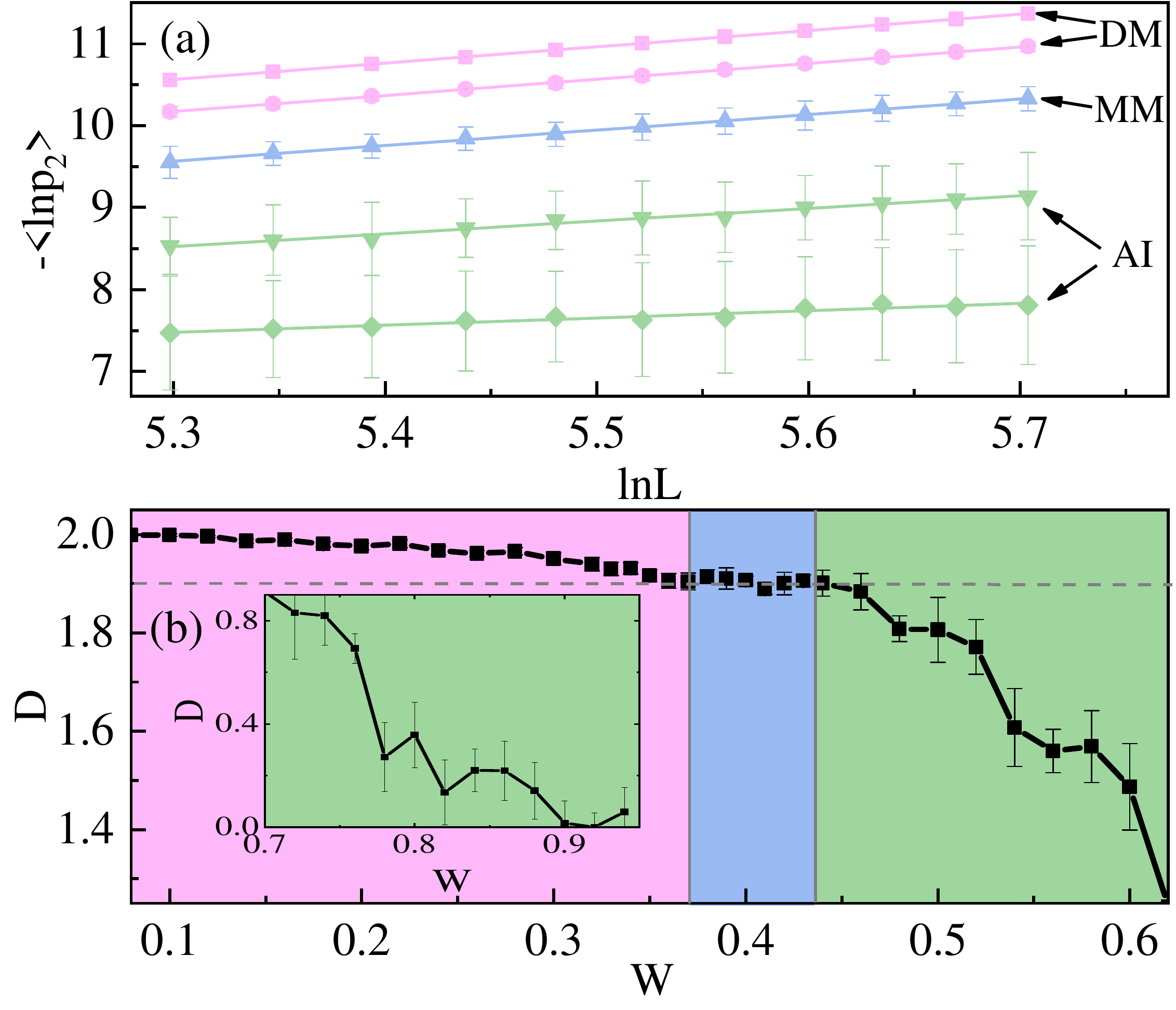}
\caption{(a) $-\langle \mathrm{ln}p_{2}\rangle$ vs $\mathrm{ln}L$
for $E=0.2$. Several disorders in different phases are
chosen. DM: $W=0.08$ (squares) and 0.28 (circles);
MM: $W=0.42$ (up-triangles);
AI: $W=0.56$ (down-triangles) and 0.66 (diamonds).
Solid lines are linear fits of numerical data.
(b) $D$ as a function of $W$ for $E=0.2$ (squares).
Dash line shows the plateau of $D=1.90$. The three
phases, colored by magenta (DM), blue (MM), and
green (AI), are identified according to
data in Fig.~\ref{fig2}.}
\label{fig3}
\end{figure}

Numerically, we use the exact diagonalization to find
the eigenfunctions of model~\eqref{hamiltonian1}. In
our scenario, we construct the tight-binding Hamiltonian
by the Kwant package \cite{gwgroth1}
in Python and solve the eigenequation
$H\psi=E\psi$ by the Scipy library \cite{scipy} for $L$ varying from
200 to 300 . The average logarithms of IPR as a function
of $\ln L$ for $W=0.08$, 0.28, 0.42, 0.56, and 0.66 at
$E=0.2$ (the same as Fig.~\ref{fig2}(a)) are shown
in Fig.~\ref{fig3}(a). The corresponding curves are virtually
straight lines, which provide strong evidence for the scaling law
Eq.~\eqref{ipr_scaling}. The slopes (fractal dimensions)
clearly decrease with the increasing of $W$ from the DM phase
to the AI phase, and $D=1.90\pm0.02$
for $W_{c,1}<W=0.42<W_{c,2}$ in the MM phase.
\par

We further authenticate the universality of the
fractal nature by displaying $D(W)$ at $E=0.2$
for a large range of disorders covering the
three phases in Fig.~\ref{fig3}(b). Apparently,
a plateau of $D=1.90$ is observed in the MM phase
determined by data in Fig.~\ref{fig2}(a),
indicating that the fractal dimension of the
fixed line does not depend on $W$.
For $W<W_{c,1}$ (DM), wave functions are not a fractal
any more since $D\simeq d$, while, for $W\gg W_{c,2}$,
IPRs are found to be independent of $L$, i.e.,
$D=0$, see the inset of Fig.~\ref{fig3}(b),
a typical feature for AIs. Noticeably,
wave functions near MM-AI transitions and on the
localized side, for example $W\in[0.44,0.64]>W_{c,2}$, have
fractal structures as well, which can be contributed
to the finite-size effect, rather than the criticality,
since $D$ will decrease if we evaluate it by using
larger system sizes \cite{supp}.
\par

\begin{figure}[ht!]
\includegraphics[width=0.45\textwidth]{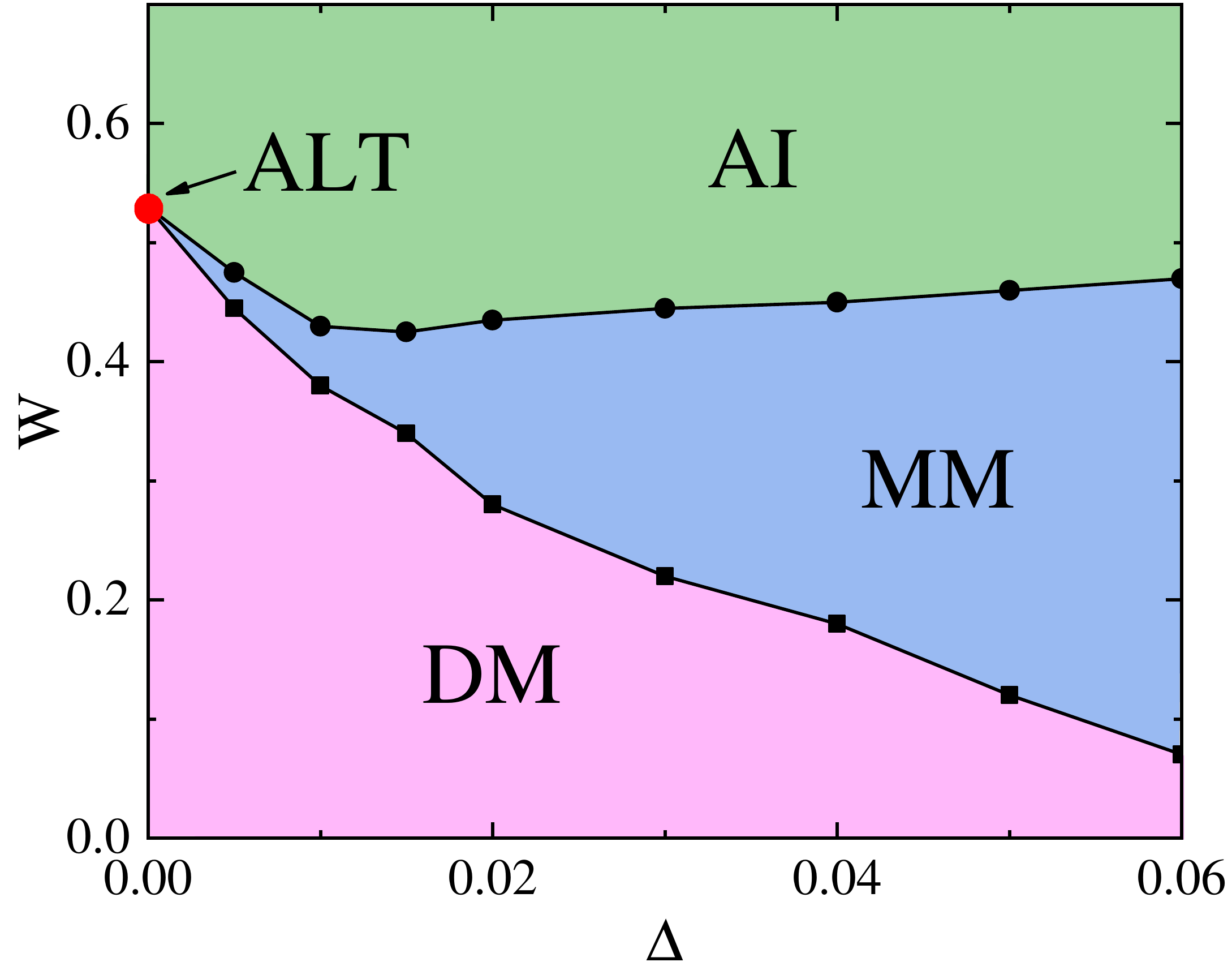}
\caption{Schematic phase diagram in the $\Delta-W$ plane:
DM (magenta), MM (blue), and AI (green) at $E=0.2$.
Only an isolated critical
level exists at $\Delta=0$ (symplectic ensemble) with
$\xi\propto|W-W_c|^{-\nu}$ and $\nu=2.8$. For $\Delta\neq 0$,
the MM phase exists within a window of
$[W_{c,1}(\Delta),W_{c,2}(\Delta)]$, and $\xi$ in the vicinity
of $W_{c,1}$ (black squares) and $W_{c,2}$ (black circles) diverges
exponentially as $\xi\propto\exp[\alpha/\sqrt{|W-W_c|}]$.}
\label{fig4}
\end{figure}

It is also enlightening to compare the fractal dimensions
of the MM phase in model~\eqref{hamiltonian1}
with those of critical states in other 2D materials.
The fractal dimension of isolated critical levels for ALTs
in the symplectic ensemble is found to be $1.66\pm 0.05$
\cite{hobuse1} while $D=1.75$ for the quantum Hall type
criticality \cite{fevers2}.
Thus, wave functions in the MM phase with $D=1.90\pm0.02$
occupy a larger space than those critical states of random
SU(2) model under strong magnetic fields \cite{cwang2}.
\par

A more inclusive picture is procured by executing
exhaustive simulations for different $\Delta$ at $E=0.2$.
The fixed line persists at finite $\Delta$
as expected such that the three phases coexist,
and model~\eqref{hamiltonian1} always experiences the
DM-MM-AI phase transitions at $\Delta-$dependent
transition points $W_{c,1}$ and $W_{c,2}$. While,
for $\Delta=0$, there is only one quantum critical
state at which the system undergoes a normal ALT.
Furthermore, it is numerically justified that the correlation
lengths $\xi$ always diverge as $\xi\propto\exp
[\alpha/\sqrt{|W-W_{c}|}]$ near $W_{c,1}$
(squares) and $W_{c,2}$ (circles) in Fig.~\ref{fig4},
see clarifications in Supplemental Materials \cite{supp}.
\par

In conclusion, analyses of the dimensionless conductance and
the IPR provide substantial evidence to the existence of an
unusual marginal metallic phase between the diffusive metal
and the Anderson insulator in ferromagnetic 2DEGs with SOIs.
Such systems undergo a DM-MM-AI transition as either disorder strength
or Fermi level varies. Near the transition points, the conductance
can be described well by the one-parameter scaling hypothesis.
The criticality of the DM-MM-AI transitions is consistent with
universality description of a quantum phase transition in the
sense that correlation lengths diverge as an exponential of an
inverse square root of $|W-W_c|$ for all critical points.
Besides, eigenfunctions in the MM phase are of
fractals of dimension $D=1.90$, while extended states
in the DM phase spread over the entire lattice.
A schematic phase diagram in the $\Delta-W$ plane is presented.
\par

\begin{acknowledgments}
This work is supported by the National Key Research \& Development
Program of China (Grants No. 2016YFA0200604),
the National Natural Science Foundation of China (Grants No.~11774296,
11704061, 21873088, and 11874337), and Hong Kong RGC (Grants No.~16301518 and 16300117).
C.W. is supported by UESTC and the China Postdoctoral Science Foundation
(Grants No.~2017M610595 and 2017T100684). C.W. also acknowledges
the support from Peng Yan and Xiansi Wang.
\par
\end{acknowledgments}


\begin{thebibliography}{99}
\bibitem{note_beta} We cannot use the numerical data at hand to depict
$\beta(g_L)$ for large $g_L$. Instead, at this stage, we use an analytical
formula $\beta(g)=f_1 g^{-1}+f_4 g^{-4}+o(g^{-5})$ based on the
non-linear $\sigma$ model \cite{shikami1} with $f_1$ and $f_4$ being fitting parameters.
The exhaustive illustration of $\beta(g)$ in the large $g$ limit surely
deserves further efforts, which, however, is not the main focus in this
work.
\end{thebibliography}
\end{document}